\documentclass[mypaper,7pt,twoside]{CoAst}
\usepackage{epsf,graphicx,fancyhdr}
\renewcommand{\chaptermark}[1]{\markboth{#1}{}}

\newcommand{\kopf}{\small\itshape Comm. in Asteroseismology \\ Vol. 53, 2008}

\newcommand{\Authors}[1]{\begin{center}\normalsize\bf\sf #1 \end{center}}

\renewcommand{\author}[1]{\begin{center}\normalsize\bf\sf #1 \end{center}}
\newcommand{\Address}[1]{\begin{center}\small\sf #1 \end{center}}

\newcommand{\Objects}[1]{\begin{flushleft}{\hspace{0cm}\small Individual Object: }\small\sf #1 \end{flushleft}}

\renewenvironment{abstract}{\section*{Abstract}\normalsize\sf}{}
\newcommand{\References}[1]{\begin{flushleft}{\large References\\}\vspace*{2mm}\small #1 \end{flushleft}}

\newcommand{\chapterDSSN}[2]{\chapter[\sf\normalsize #1\\ \footnotesize \hspace*{5mm}by #2 \sf\normalsize][]{#1\\}\rhead[\fancyplain{}{\sf\footnotesize \center{#1}}]{\fancyplain{}{\sffamily\thepage}}\lhead[\fancyplain{\kopf}{\sffamily\thepage}]{\fancyplain{\kopf}{\sf\footnotesize \center{#2}}}}

\renewcommand{\chaptername}{}

\newcommand{\acknowledgments}[1]{\vspace*{5mm}\noindent  \textbf{Acknowledgments.} #1}

\def\rfr{\smallskip\par\noindent
        \hangindent=7truemm
        \hangafter=1}

\begin{document}
\sf

\chapterDSSN{CCD Photometry of $\delta$ Scuti stars 7 Aql and 8 Aql}{L. Fox Machado,  R. Michel, M. \'Alvarez, L. Parrao,  
A. Castro, J.H. Pe\~na}

\Authors{L. Fox Machado$^1$, R. Michel$^1$, M. \'Alvarez$^1$, L. Parrao$^2$, A. Castro$^1$,  J. H. Pe\~na$^2$} 
\Address{$^1$
Observatorio Astron\'omico Nacional, Instituto de Astronom\'{\i}a,
Universidad Nacional Aut\'onoma de M\'exico, Ensenada B.C., Apdo.
Postal 877,
M\'exico\\
$^2$ Instituto de Astronom\'{\i}a, Universidad Nacional Aut\'onoma
de M\'exico, M\'exico D.F., Apdo. Postal 70-264,
M\'exico\\}

\noindent
\begin{abstract}
As a continuation of the study
of the $\delta$ Scuti stars 7 Aql and 8 Aql; new CCD photometric
data were acquired in 2007. We present a period analysis on these
data that confirm the dominant modes detected in each star in the
framework of the STEPHI XII campaign in 2003.

\end{abstract}

\Objects{7 Aql; 8 Aql; GSC 05118-00503; GSC 05118-00553; GSC 05118-00513}

\section{Introduction}

$\delta$ Scuti variables are promising candidates for helping in
understanding the internal processes occurring in the interiors of
intermediate mass stars. Most of the $\delta$ Scuti stars pulsate,
simultaneously, in a number of radial and non-radial modes  showing
complicated pulsation spectra. Since long time series are required
to disentangle their pulsation spectra,
 multisite coordinated campaigns are often undertaken
(e.g., STEPHI [Stellar Photometry International; Michel et al. 2000] or
 Delta Scuti Network; Breger et al. 1999])
and space observations are underway (COROT [Baglin 2003; Michel et al. 2005]).

\bigskip
7~Aql (HD~174532, SAO~142696, HIP~92501) is a $\delta$~Scuti
variable discovered in a systematic search and characterization of
new variables in preparation for the COROT mission (Poretti et al.
2003) and was selected as the main target of the STEPHI XII
multisite campaign in 2003. In this campaign 8~Aql (HD~174589,
SAO~142706, HIP~92524) was used as the only comparison star because
there are no other bright stars in the field-of-view (FOV) of the
4-channel photometer used in the STEPHI network.

 Nonetheless, by carefully analyzing the derived
differential light curve 7 Aql - 8 Aql  and individual
nondifferential light curves  we were able to demonstrate that 8 Aql
is a new $\delta$ Scuti variable. Moreover, it was shown that the
amplitude spectrum of both stars are not superposed. Three and seven
frequency peaks were unambiguously detected with a 99 \% confidence
level  in 8 Aql and 7 Aql respectively and a possible identification
of the observed modes in terms of radial order was performed (Fox
Machado et al. 2007).

\bigskip
As a continuation of this study we have carried out CCD photometric
observations in 2007 of 7 Aql and 8 Aql to obtain differential
photometry of these stars with respect to much dimmer comparison
star. The primary results of these observations are present in this
paper.

\begin{table}[!t]\centering
  \setlength{\tabcolsep}{1.0\tabcolsep}
 \caption{Positions, magnitudes and spectral type of targets and selected comparison stars.}
  \begin{tabular}{llcrcl}
\hline
Star&  $ID$ & RA & Dec & V  &  SpTyp \\
&&(2000.0)&(2000.0)&(mag)&\\
\hline
7 Aql &HD 174532& 18 51 05   &-03 15 40.1 &6.9  & A2   \\
8 Aql & HD 174589& 18 51 22   &-03 19 04.2 &6.1  &  F2   \\
 C1 & GSC 05118-00503& 18 51 09    &-03 18 52.2  & 12.6 &- \\
 C2 &GSC 05118-00553& 18 51 04    &-03 18 36.1  & 13.6 & - \\
 C3 &GSC 05118-00513& 18 51 02    &-03 17 32.5  & 12.1 & -  \\
\hline
\end{tabular}
\end{table}

\section{Observations and data reduction}
CCD photometric observations of 7 Aql and 8 Aql were carried out for
fourteen nights (about 86 hours of data)  over the period 2007 June
14-21 and July 7-12 at the Observatorio Astr\'onomico Nacional, in
San Pedro M\'artir, Baja California,  M\'exico. The observations
were performed using the 0.84 m  telescope to which a CCD camera was
attached. A  $1024 \times 1024$ CCD camera  was used in June with a
plate scale of $0.38''$/pixel, while
 a $2048 \times 2048$  camera was implemented in July which
has a plate scale of $0.15''$/pixel. The observations  were obtained
through a Johnson $V$ filter.

\begin{figure}[!t]
%\begin{center}
\includegraphics[width=8.cm]{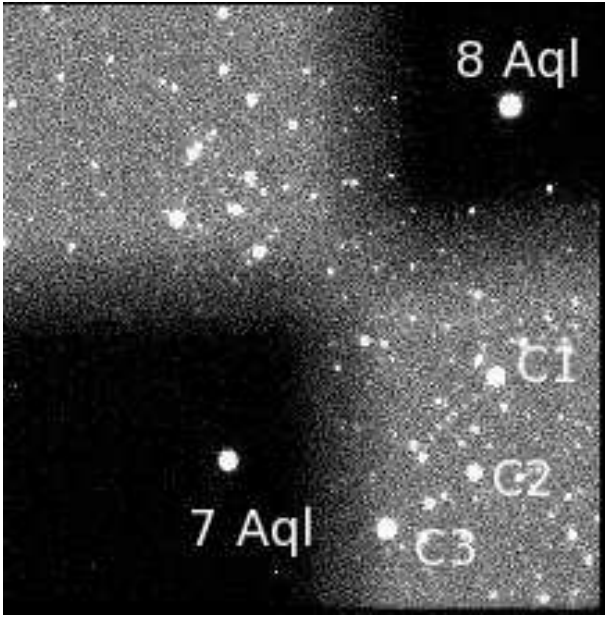} \caption{Image of the CCD field-of-view ($7.5'
\times 7.5'$). The reference stars are marked as C1, C2 and C3.
North is down and East is right.}
%\end{center}
\end{figure}

\bigskip
Figure 1 shows a typical image of the CCD's  FOV ($7.5' \times 7.5'$
at the f/15 focus of the telescope) during the runs. This FOV
supplies a large set of reference stars for the two variables. We
have selected, however,  the brighter ones  after 7 Aql and 8 Aql
for computing differential magnitudes. The coordinates, $V$
magnitudes and identifications
 of the reference stars (marked as C1, C2 and C3 in Figure 1) and
main targets  are listed in Table 1. Pieces of gelatin neutral
density filter (Kodak Wratten ND 2.0) with a transmission of $\sim$
1\%  placed on top of the V filter (the dark regions in Figure 1)
were used to reduce the brightness of the variable stars at the
detector since, as can be seen in Table 1, they are much brighter
than the selected comparison stars in the CCD's FOV. This allowed us
to obtain images with high signal-to-noise ($S/N$) ratios in both
the pulsating and comparison stars with an exposure time  of about
90 s without reaching the saturation levels of the CCD detectors.
The acquired images were reduced in the standard way using the IRAF
package. Aperture photometry was implemented to extract the
instrumental magnitudes of the stars. All differential magnitudes
have been computed using GSC 05118-00503 (C1) as the principal
comparison star and GSC 05118-00553 (C2) as the check star. GSC
05118-00513  (C3) was not useful neither as comparison nor as check
star because, as can be seen in Figure 1, in some CCD images it was
so close to  the neutral density filter that its magnitude was
affected. The differential magnitudes were normalized by subtracting
the mean of differential magnitudes for each night.

\bigskip
 The differential light curves 8 Aql - C1, 7 Aql - C1 for
three selected nights are illustrated in Figure 2 (the three top and three middle 
plots respectively). As can be seen,
the oscillations of 8 Aql and 7 Aql are clearly inferred, with the
dominant period of 8 Aql longer than that of 7 Aql. The magnitude
differences between comparison stars C1 - C2  were also derived in
order to confirm their constancy. As can be seen in
Figure 2 (the three bottom plots), no indications of photometric
variability on these stars was found.

\begin{figure}[!t]
\begin{center}
\includegraphics[width=8.0cm]{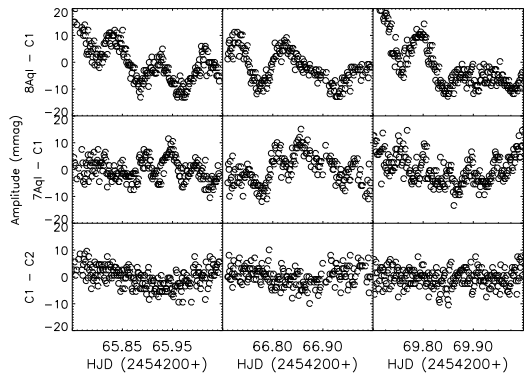}
\caption{Examples of the differential light curves with reference
star 'C1'. The three top and middle plots correspond to differential light
curve 8 Aql - C1 and 7 Aql - C1 respectively. The three bottom plots are for  
C1 - C2.}
\end{center}
\end{figure}

\section{Frequency analysis}
The amplitude spectra of the differential time series were obtained
by means of an iterative sinus wave fit (ISWF; Ponman 1981) and  the
software package Period04 (Lenz \& Breger 2005). In both cases the
frequency peaks are obtained by applying a non-linear fit to the
data. Since both packages yielded similar results, we present the
spectral analysis only in terms of ISWF. The window function of the
observations is shown in Figure 3. As is known, observations from a
single site yield a worse window function as compared to 
multisite observations, hence the sidelobes in
the spectral window are at 90\% of the main lobe. Even so, the
resulting amplitude spectra are good enough to determine the main
pulsation modes in each star.

\begin{figure}[!t]
\begin{center}
\includegraphics[width=9.1cm]{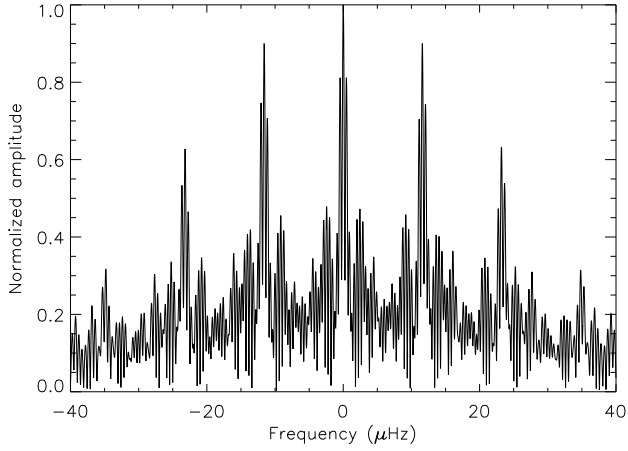}
\caption{Spectral window in amplitude. The first sidelobes are at 90
\% of the main lobe.}
\end{center}
\end{figure}

\begin{figure}[!hb]
\begin{center}
\includegraphics[width=9.1cm]{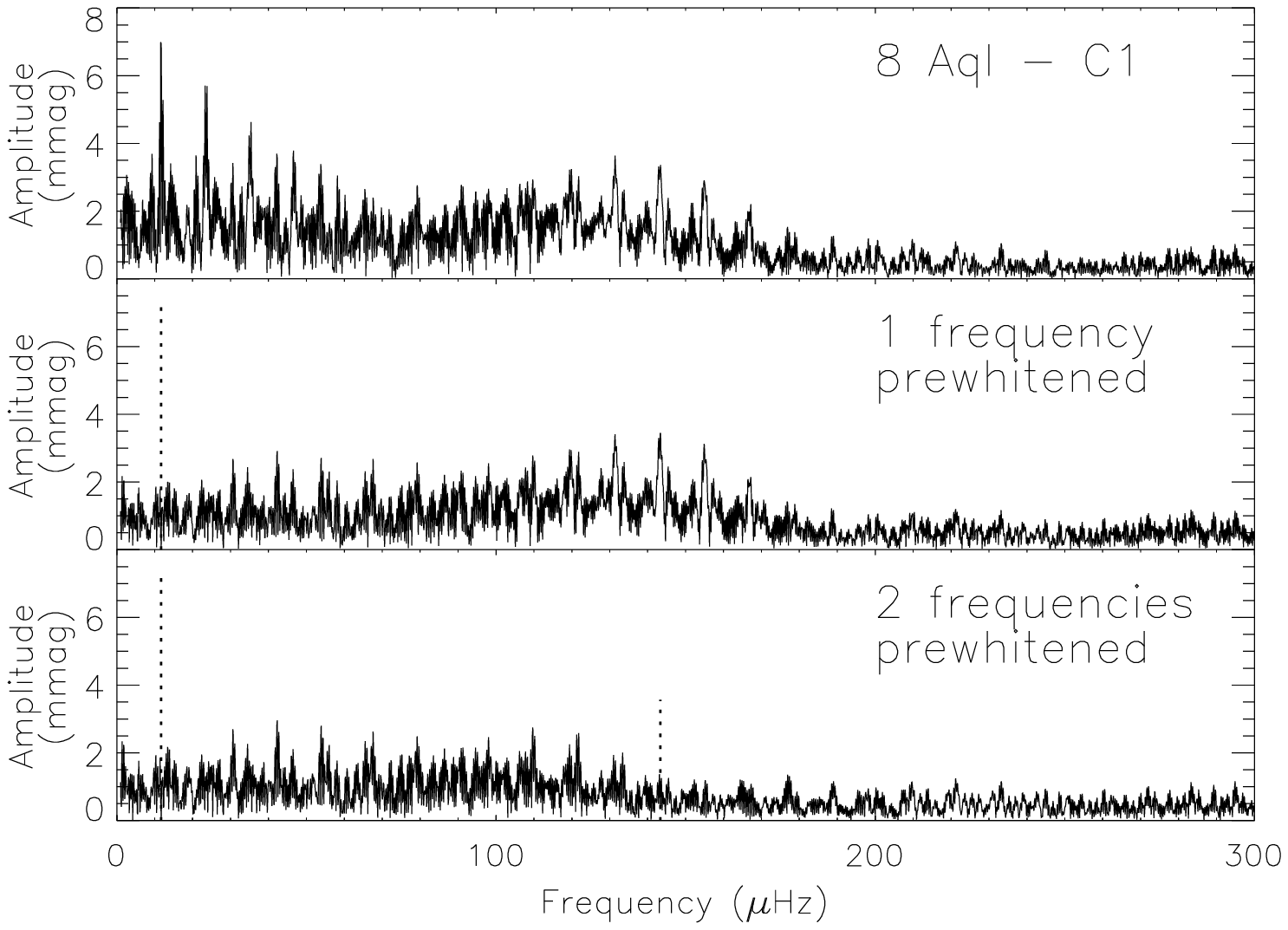}
\caption{Amplitude spectrum  8 Aql - C1 and the prewhitening process
of the detected peaks. }
\end{center}
\end{figure}

\begin{figure}[!t]
\begin{center}
\includegraphics[width=9.1cm]{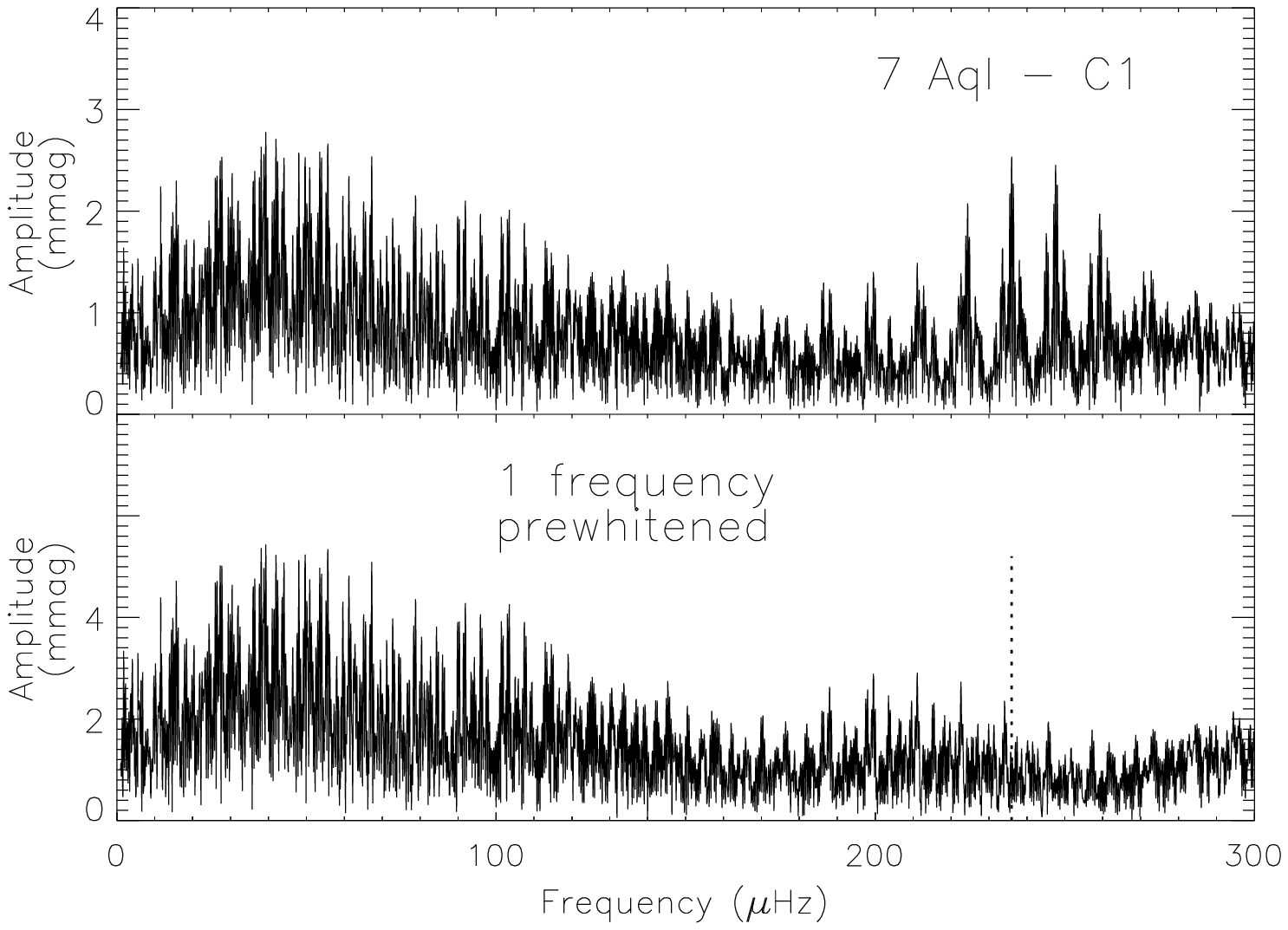}
\caption{Amplitude spectrum 7 Aql - C1 and the prewhitening process
of the detected peaks. }
\end{center}
\end{figure}

\begin{figure}[!b]
\begin{center}
\includegraphics[width=9.1cm]{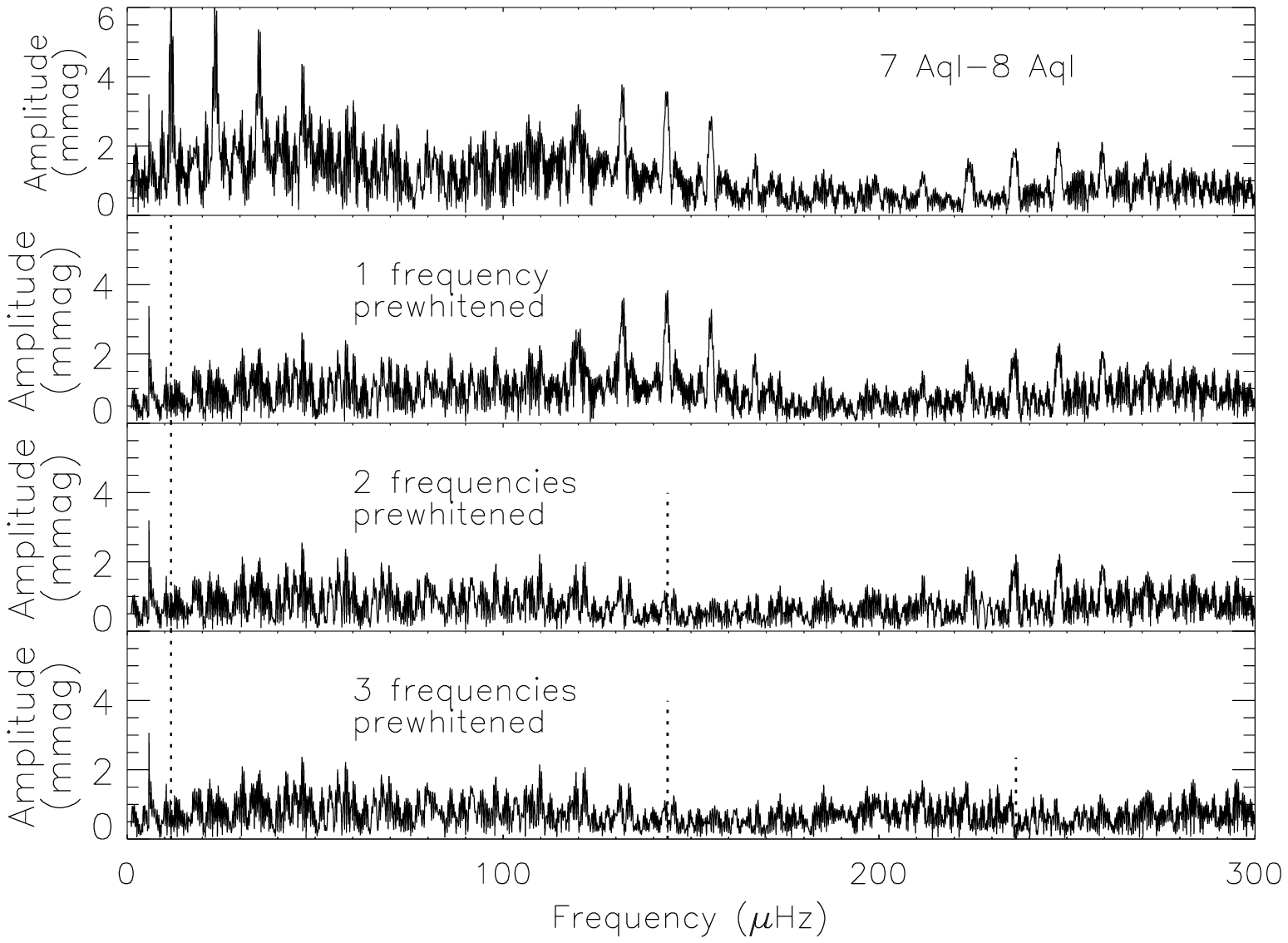}
\caption{ Amplitude spectrum 7 Aql - 8 Aql and the prewhitening
process of the detected peaks.}
\end{center}
\end{figure}

\begin{figure}[!t]
\begin{center}
\includegraphics[width=9.1cm]{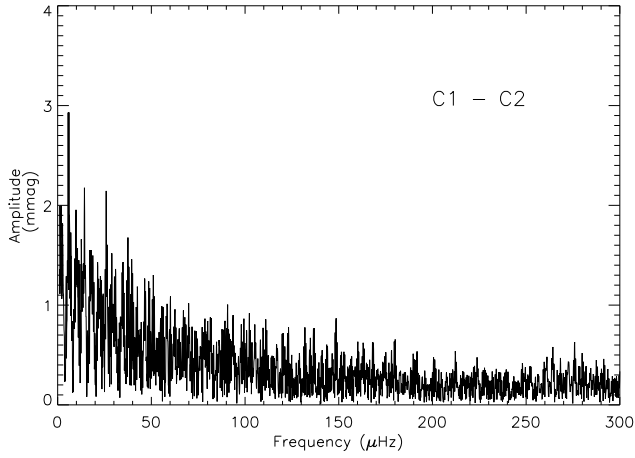}
\caption{Amplitude spectrum  C1 - C2. }
\end{center}
\end{figure}

\begin{table}[!h]
\begin{center}
\caption{Detected frequencies in the present study and the dominant
modes detected in STEPHI XII campaign. $\nu_{\rm a}$ is the first
harmonic of the day.}
\begin{tabular}{lrcrlccr}
\noalign{\smallskip} \hline \noalign{\smallskip}
\multispan4 \hfill This paper (2007) \hfill &\multispan4 \hfill STEPHI XII (2003) \hfill \\
\hline \hline
&&&  \vrule &&&&\\
  & $\nu$  & A  & $S/N$ \vrule& & $\nu$  & A &  $S/N$ \\
      & ($\mu$Hz)& (mmag)& \vrule & &($\mu$Hz) &(mmag) & \\
\hline
&&&&&&&\\
\multispan4 \hfill 8 Aql - C1  \hfill &\multispan4 \hfill 8 Aql \hfill \\
\hline
$\nu_{\rm a}$& 11.69 & 7.4  &- \vrule&$\nu_{\rm a}$&-&-&- \\
$\nu_{1}$  &143.36 & 3.6&3.0 \vrule &$\nu_{1}$& 143.36&9.8& 7.9\\
\hline
&&&&&&&\\
\multispan4 \hfill 7 Aql - C1  \hfill &\multispan4 \hfill 7 Aql \hfill \\
\hline
$\nu_{1}$&235.96  &2.6 &3.8 \vrule &$\nu_{1}$&236.44&5.8&5.5 \\
\hline
&&&&&&&\\
\multispan4 \hfill 7 Aql - 8 Aql  \hfill &\multispan4 \hfill 7 Aql - 8 Aql \hfill \\
\hline
$\nu_{\rm a}$&11.69 &9.2& -  \vrule &$\nu_{\rm a}$& -&-&- \\
$\nu_{2}$&143.76&4.0&3.7 \vrule & $\nu_{2}$&143.36&9.8&15.7\\
$\nu_{3}$&236.45&2.4&3.0 \vrule & $\nu_{3}$&236.44&9.6&9.9\\
\hline
\end{tabular}
\normalsize
\end{center}
\label{tab:dom}
\end{table}

\bigskip
The amplitude spectra  of the differential light curves 8 Aql$-$C1
and 7 Aql$-$C1  are shown in the top panel of Figure 4 and Figure 5,
respectively.  We note that below 100 $\mu$Hz these spectra are
dominated by the daily aliasing. A quick
inspection of these spectra in the region $> 100\, \mu$Hz shows that
the amplitude spectrum of 8 Aql present the main peaks
 between 100 and 200 $\mu$Hz, while in the amplitude spectrum of 7 Aql the peaks
are between  200 and 300 $\mu$Hz. The subsequent panels in the
figures, from top to bottom, illustrate the prewhitening process of
the frequency peaks in each amplitude spectrum. We followed the same
procedure as explained in \'Alvarez et al. 1998.
 In the case of 8 Aql  
(Fig. 4) the  highest amplitude peak is located at 143.36 $\mu$Hz,
but after prewhitening the
 peak at 11.69 $\mu$Hz. This latter is 
most likely to be the first harmonic of the day and it is not considered hereafter. 
On the other hand, 7 Aql (Fig. 5) shows a frequency peak at 235.96
$\mu$Hz. Both  frequency peaks are confirmed in the amplitude
spectrum of 8 Aql - 7 Aql (Fig. 6) after prewhitening the peak
at 11.69 $\mu$Hz.  The remaining peaks which seem to be weakly present in
all amplitude spectra might be disregarded because they do not
significantly improve the residuals. These peaks are
indistinguishable due to the aliasing caused by the window function.
In any case our goal was to detect the main pulsation modes in each
star. Figure 7 illustrates the amplitude spectra C1 - C2. As can be
seen the typical $1/f$ noise dominates whole spectrum. It confirms
the non-oscillatory behavior in both stars.

\section{Discussion}
The frequencies, amplitudes and $S/N$ ratio in amplitude of the
peaks detected in the amplitude spectra 7 Aql $-$ C1 (Fig. 4), 8 Aql
$-$ C1 (Fig. 5) and 7 Aql$-$8 Aql (Fig. 6) are summarized in Table
2. For comparison also shown are the dominant modes detected in
nondifferential and differential time series derived in STEPHI XII
campaign. It is evident that the dominant modes detected in each
star in STEPHI XII campaign are confirmed. There is a fairly good
agreement between the frequency values derived in this survey and
those derived from multisite photoelectric photometry, besides that
onesite observations are less accurate than the multisite ones. The
disagreement in the amplitudes can be attributed to the much slower
sampling rate of the CCD detectors.  In fact, this might have
suppressed not only the amplitude of the main oscillation modes but
also the secondary harmonics found in STEPHI XII campaign.

\section{Conclusions}
We have presented the analysis of new CCD photometric observations
of $\delta$ Scuti stars 7 Aql and 8 Aql carried out during fourteen
nights in June and July, 2007 at the Observatorio Astron\'omico
Nacional, M\'exico. About 86 hours of time resolved CCD differential
photometry  were obtained. The dominant oscillation frequencies
detected in 2003 in the framework of STEPHI XII campaign by means of
nondifferential photometry have been confirmed in this season by
using CCD differential photometry.

\acknowledgments{
We would like thank the assistance of the staff of
the OAN during the observations. This paper was partially supported
by Papiit IN108106.
}

\References{
\rfr Alvarez, M.,  et al., 1998, A\&A 340, 149\\
\rfr Baglin, A., Adv. Space Res., 2003, 32, 345\\
\rfr Breger, M., et al., 1999, A\&A 349, 225\\
\rfr Fox Machado, et al., 2007, AJ 134, 860\\
\rfr Lenz, P. \& Breger, M., 2005, CoAst 146, 53\\
\rfr Michel, E., et al., 2000, ASP Conf. Ser., 203, 483\\
\rfr Michel, E., et al., 2005, ASP Conf. Ser., 333, 265\\
\rfr Ponman, T., 1981, MNRAS, 196, 583\\
\rfr Poretti, E., et al., 2003, A\&A 406, 203\\
}

\end{document}